\documentclass[aps,prb,twocolumn,10pt]{revtex4-1}
\usepackage{ifpdf}
\usepackage{amsmath}
\usepackage{subfig}
\usepackage{graphicx}
\usepackage{epstopdf}
\usepackage{mathtools}
\usepackage{color}
\usepackage{float}
\begin{document}

\title{Phonon Induced Instabilities in Correlated Electron Hamiltonians}
\author{Nahom K. Yirga}
\affiliation{Department of Physics, Boston University, Boston, Massachusetts 02215, USA}

\author{Ka-Ming Tam}
\affiliation{Department of Physics and Astronomy, Louisiana State University, Baton Rouge, Louisiana 70803, USA}
\affiliation{Center for Computation and Technology, Louisiana State University, Baton Rouge, LA 70803, USA}

\author{David K. Campbell}
\affiliation{Department of Physics, Boston University, Boston, Massachusetts 02215, USA }
\date{\today}

\begin{abstract}
Studies of Hamiltonians modeling the coupling between electrons as well as to local phonon excitations have been fundamental in capturing the novel ordering seen in many quasi-one dimensional condensed matter systems. Extending studies of such Hamiltonians to quasi-two dimensional systems is of great current interest, as electron-phonon couplings are predicted to play a major role in the stabilization or enhancement of novel phases in 2D material systems. In this work, we study model systems that describe the interplay between the Hubbard coupling and the phonon modes in the Holstein (H) and Su-Schrieffer-Heeger (SSH) Hamiltonians using the functional renormalization group (fRG). For both types of electron phonon couplings, we find the predicted charge density wave phases in competition with anti-ferromagnetic ($AF$) ordering. As the system is doped, the transition shifts, with both orders showing incommensurate peaks. We compare the evolution of the quasiparticle weight for the Holstein model with that of the SSH model as the systems transition from antiferromagnetic to charge-ordered ground states. Finally, we calculate the self-energy of the phonon and capture the impact of charge ordering on the phonon modes.
\end{abstract}
\maketitle

\section{Introduction}
Many of the novel phenomena observed in low dimensional electronic systems are driven by the combined effects of electron-electron (e-e) and electron-phonon (e-ph) interactions.  Interactions among electrons drive charge and spin fluctuations which can lead to ordering of the spin and charge densities with the remnants of the density order in the doped system serving to stabilize various types of superconducting order. Electron-phonon interactions can dramatically modify these orders by distorting the electronic band structure, altering the mobility in conductors and providing the mechanism for conventional superconductivity. The interplay between these two interactions has helped explain the physics of conducting polymers\cite{baeriswyl2012conjugated,baeriswyl1992overview}, superconducting order in the fullerenes\cite{koch1999screening} and density wave orders in the charge transfer solids\cite{clay2003pattern,clay2019charge}. But even in systems in which the leading order is driven primarily by only one of these interactions, the impact of the other interaction can be significant. The high $T_c$ Cuprates are a prime example, with the phonon modes predicted to have considerable impact on the physics of the material occuring at frequencies far below the hopping and coupled weakly in contrast to the strong e-e interactions in the system \cite{devereaux2004anisotropic,johnston2010systematic,lanzara2001evidence,da2014ubiquitous}. Similarly, recent studies of superconductivity in FeSe heterostructures show an up-to-an-order of magnitude enhancement in the critical temperature, much of it attributed to the coupling of electrons to phonons in the substrate \cite{zhang2017ubiquitous,xu2020superconductivity}. The interplay between these interactions also explains the charge ordering observed in competition with superconductivity in the transition metal dichalcogenides\cite{neto2001charge,chen2016charge,manzeli20172d}.

Beyond the novel orderings due to the interplay between these interactions, an accounting of the couplings is necessary for a quantitative description of the materials. An excellent example of this are models of conducting polymers, which require Hubbard-like e-e couplings along with the dominant Su-Schrieffer-Heeger (SSH) e-ph interactions in order to explain the optical absorption spectra observed in these systems \cite{baeriswyl2012conjugated}. In the resulting SSH-Hubbard (SSHH) models we see the standard transition from a Peierls phase, a bond ordered density wave ($BOW$), to an antiferromagnet ($AF$) with the critical e-ph coupling at the transition decreasing to zero with phonon frequency\cite{sengupta2003peierls}. Studies of the model found that bond correlations are  {\it enhanced} by Coulomb interactions up to intermediate values of order the bandwidth, with retardation effects not playing much of a role\cite{campbell1988bond,weber2015excitation}. In general systems, the distinction between inter-site (SSH) electron-phonon coupling and intra-site (Holstein or Molecular Crystal) e-ph coupling, along with the frequency and dispersion relation of the phonons, can lead to important differences, as was demonstrated in the pioneering papers of Fradkin and Hirsch\cite{hirsch1983phase,fradkin1983phase}.

\begin{figure*}
\centering
\includegraphics[scale=0.59]{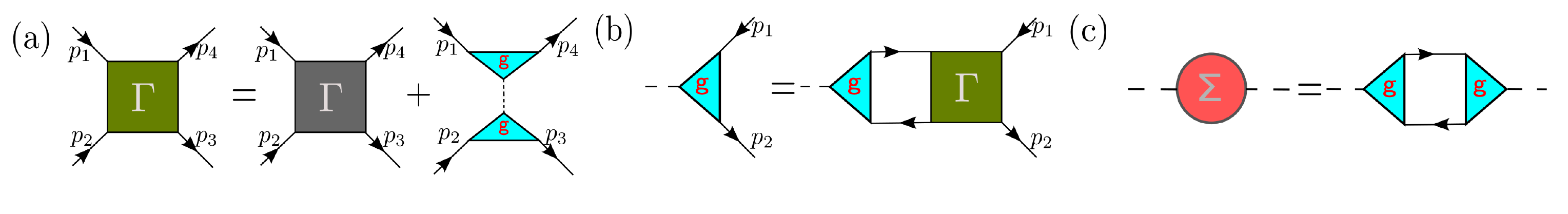}
\caption{Equations for the electron and phonon vertices. (a) The effective dynamic vertex constructed from the electron-electron and electron-phonon vertices. (b) Equation for the electron-phonon vertex in terms of the dynamic vertex given in (a). (c) Expression for the self energy of the phonon }
\end{figure*}

A number of studies have explored the nature of possible phases in two-dimensional models involving e-ph and e-e interactions. Studies of the 2D Hubbard-Holstein (HH) Hamiltonian show competition between an antiferromagnet ($AF$) and a charge density wave ($CDW$) with a possible metallic phase at the critical transition line. The existence and size of the finite region of metallicity in the 2D HH model has been difficult to determine as variational Monte Carlo studies find a metallic phase whose domain size is inconsistent for different phonon frequencies\cite{ohgoe2017competition,karakuzu2017superconductivity} whereas the correlated metallic phase captured in determinant Monte Carlo studies at phonon frequencies of order the hopping does not extend to the Holstein model at $U=0$\cite{johnston2013determinant,costa2020phase}. Earlier Quantum Monte Carlo (QMC) results suggested the $CDW$ phase as the likely sole ground state of the 2D Holstein and Hubbard-Holstein models with the difference to the one dimensional system possibly due to the larger electronic susceptibility\cite{costa2020phase} and weak coupling $CDW$ instability\cite{hohenadler2019dominant,weber2018two}. Before finite size scaling analysis, the transition to a metallic phases is found at $\lambda_c=0.61t$\cite{costa2020phase} while the sensitivity of the metallic phase to phonon frequencies larger than the hopping remains an open question. The qualitatively similar phase diagram for the model in 1D shows the domain of metallicity expanding with phonon frequency, indicating a robust competition between the two interactions, and an extrapolation of this behavior to the 2D system suggests a metallic phase in the 2D Holstein model\cite{costa2021magnetism,tezuka2007phase}. Alternatively, studies of the 2D SSH and SSHH models show a first-order transition from an $AF$ to a $BOW$ phase occurring at a finite e-ph coupling for all phonon frequencies\cite{xing2021quantum,cai2021antiferromagnetism,feng2021phase}.The transition shows only minor changes as the local Hubbard coupling is altered, suggesting little competition between the two interactions\cite{feng2021phase}. Studies of the SSH phonon carried out with a focus on polyacetylene found stronger competition between the $AF$ and $BOW$ phases, but modeling polyacetylene requires the additional nearest neighbor density-density coupling ($V$) to obtain agreement with the optical absorption spectra, which has the additional effect of enhancing $BOW$ ordering in 1D systems\cite{sengupta2003peierls,gallagher1997excitons}. Finally, there have been limited studies of the impact of doping on these systems with most results confined to 1D. DMRG studies of the doped HH model show little change due to doping, with superconducting fluctuations on a par with charge fluctuations for much of the doping regime around half filling\cite{tezuka2007phase}.

In the present work, we address a number of open questions in these systems by analyzing the impact of doping and the phonon frequency in the HH and SSHH models in one and two dimensions. The models incorporate the on-site Hubbard coupling ($U$) and the nearest neighbor density-density interaction ($V$), along with a coupling to a phonon mode which can be either of the SSH type or the Holstein type. We present a functional renormalization group (fRG) study of the system from two perspectives. First, we integrate out the quadratic phonon fields and run the flow for the fermions with a new effective two-particle vertex. As the displacements in the lattice are coupled to the electrons (the density operator in the case of the Holstein phonon and the hopping operator for the SSH phonon), the general e-ph coupling has a non-trivial momentum structure. Integrating out the phonons couples these e-ph vertices, leading to a dynamical e-e vertex. Within this picture, the fRG captures the transition from an $AF$ phase to a charge-ordered phase as a function of the e-ph coupling. Further access to the electron self-energy shows the deformation due to the phonons with asymmetries from the modes emerging in the quasiparticle weight as a function of the e-ph coupling. We study the consequences of doping and changes to the phonon frequency on this transition. Second, we study the flow of the phonon vertices as the e-ph coupling is the primary driver of deformations in the fermion self-energy. Such studies can be of general interest, as in many systems ordering in the electronic sector can lead to a softening of phonon modes. For example, phonon softening is seen in the FeSe superconductors at the structural transition with a smaller softening as the system becomes superconducting\cite{luo2012quasiparticle}. Access to the phonon self-energy enables us to capture this softening and help quantify possible enhancements of electronic order due to phonon modes.

The remainder of the article is organized as follows. We begin in Sec.\ref{methodFRG} with the flow equations for the vertices of a general electron-phonon system. The response in the Hubbard model to a Holstein phonon is presented in Sec.\ref{holsteinFRG}. The impact of the SSH phonon on an extended Hubbard model is given in Sec.\ref{sshFRG}. Our conclusions and a summary of our results are given in Sec.\ref{summary}.

\section{The flow equations in the fRG method}
\label{methodFRG}
The functional renormalization group (fRG) has become a standard tool to study competing orders in interacting electron systems\cite{wetterich1993exact,shankar1994renormalization,metzner2012functional}. Starting from a scale-dependent action, equations for the various interaction vertices of the system are derived as functions of the scale ($\Lambda$). The flow equations track the evolution of these vertices as, scale by scale, modes are integrated out. RG methods have been crucial in the study of these models and have helped elucidate the phase diagrams of Holstein, Hubbard-Holstein and Peierls-Hubbard models in one dimension \cite{tam2007retardation,bakrim2015nature,bakrim2007quantum,tam2014dominant,tam2007phase}. In the case of electronic systems coupled to a phonon mode, there is some ambiguity as to how to regulate both propagators. Previous electron-boson fRG studies adopted a momentum regulator in both the bosonic and fermionic sectors, but as the ph-ph vertices in the system are irrelevant with the marginal and relevant vertices of the e-e and e-ph type, we choose to insert the regulator into the electron propagator\cite{schutz2005collective}. This choice simplifies the flow equations, as scale derivatives of the phonon propagator are set to zero. To capture deformations of the Fermi surface in the half-filled and doped systems we utilize a pure frequency regulator and retain all momentum modes of the lattice.

A general Hamiltonian for coupling a system of interacting electrons to a phonon mode $\lambda$ can be written as
\begin{align}
\mathcal{H}=&\sum_{k\sigma}\xi_kc_{k\sigma}^\dagger c_{k\sigma} + \sum_{k_1k_2k_3}U_{k_1,k_2,k_3,k_4}c_{k_1\uparrow}^\dagger c_{k_2\downarrow}^\dagger c_{k_3\downarrow}c_{k_4\uparrow}\nonumber\\
&+\sum_q\Omega_q^\lambda b_q^\dagger b_q + \sum_{k,q,\sigma} g_\lambda (k,q) c_{k+q,\sigma}^\dagger c_{k,\sigma}(b_q+b_{-q}^\dagger)
\end{align}
where c and b operators correspond to the electron and to the phonon modes, $\xi_k$ and $\Omega_k$ are the electron and phonon dispersions, $U_k$ is the electron-electron interaction and $g_\lambda$ represents coupling between the electron and phonon modes\cite{tsai2005renormalization}. As the phonon operators are quadratic they can be integrated out exactly, leading to an electron-electron interaction mediated by the phonon. The interaction is of the form
\begin{align}
\label{effVertex}
U_{k_1,k_2,k_3,k_4}^{eff}=U_{k_1,k_2,k_3,k_4}+g_{k_1+k_{ph},-k_{ph}}^\lambda g_{k_3,k_{ph}}^\lambda\mathcal{G}_{k_{ph}}^\lambda
\end{align}
where $k_{ph}$= ($k_3-k_2$) is the particle-hole singular mode and $\mathcal{G}^\lambda$ is the phonon propagator. The fRG flow can then be constructed identically to a pure electronic Hamiltonian with the only modification coming in as a change in the initial vertex.

Alternatively, we can construct a flow for both the electron and phonon vertices. This approach presents some difficulties, as in addition to the electron vertices we need to track the phonon self-energy and the electron-phonon vertices. As the regulator is inserted only into the fermionic propagator, the fRG equations for the phonon vertices flow with the electronic single-scale propagator. The flow for the phonon self-energy is given by
\begin{align}
\partial_\Lambda\Sigma_q^\lambda=\sum_{k,\sigma}g_{k,q}\partial_\Lambda(\mathcal{G}_k^\Lambda\mathcal{G}_{k+q}^\Lambda)g_{k+q,-q}
\label{frgPHselfenergy}
\end{align}
and the flow of the e-ph vertex is
\begin{align}
\partial_\Lambda g_{k_1,q}=&\sum_kg_{k,q}\partial_\Lambda(\mathcal{G}_{k+q}^\Lambda\mathcal{G}_k^\Lambda)g_{k+q,-q}\mathcal{G}^{\lambda}_qg_{k_1,q}+\nonumber\\
&\sum_kg_{k,q}\partial_\Lambda(\mathcal{G}_k^\Lambda\mathcal{G}_{k+q}^\Lambda)\Gamma_{k_1,-k,-k-q,k_1+q}^{eff}
\label{frgEPHvertex}
\end{align}
where $\mathcal{G}$ is the fermion propagator. The flow of the electron vertices is identical to the pure fermionic flows with the two-particle electron vertex in the equations replaced by a scale-dependent form of the effective vertex defined in Eq. \ref{effVertex}. The modification of the vertex also accounts for the contributions to the fermion self-energy from the e-ph coupling. We note that in the case of flowing phonon vertices, the effective two-particle vertex has to be constructed along each point in the flow.

Either choice leads to a system with frequency-dependent vertices. To deal with the frequency and momentum dependencies of the vertex in an efficient manner, we employ a decoupled variant of the fRG at the two-loop level\cite{yirga2021frequency}. Treatment of the frequency dependence leads to stable flows and allows us to construct the flow of the self-energy which should capture deformations from the e-ph coupling at the single particle level\cite{hille2020quantitative}. The fRG equations for the e-ph vertex and the phonon self-energy given above in Eqns.\ref{frgPHselfenergy} and \ref{frgEPHvertex} can be further simplified by using the basis expansion used to derive the decoupled fRG equations\cite{lichtenstein2017high,yirga2021frequency}. The inclusion of the e-ph vertex within the decoupled fRG framework can be achieved by expanding the fermion label in the appropriate frequency and momentum basis sets. The vertex describes a forward scattering process due to the phonon and is already parameterized by the particle-hole frequency and momentum. Thus, we utilize the same auxiliary variables used for the particle-hole channel in the decoupling of the vertex to expand the e-ph interaction\cite{yirga2021frequency}. Within the decoupled framework, the vertex $\Gamma$ is expanded in the three channels along the singular frequencies with the scaling of the vertex going from $\mathcal{O}(N_f^3N^3)$ to $\mathcal{O}(N_f N_\omega^2NN_k^2)$ with $N_f$ corresponding to the number of Matsubara frequencies retained, $N_\omega$ representing the number of frequency basis functions used for expansion, $N$ is the number of sites in the system, and $N_k$ is number of momentum basis functions. Using a similar expansion for the e-ph vertex, we have
\begin{align}
g_{m,i}^\Delta(s_{ph})=\frac{1}{N\beta}\sum_{\omega_{{ph}_x},k}g_{k,k_{ph}}(\omega,\omega_{ph}) f_m(\omega_{{ph}_x})f_i(k)
\end{align}
where $\omega$,$k$ represent the incoming frequency and momenta of the fermion and the auxiliary frequency variables $\omega_{{ph}_x}=2\omega+\omega_{ph}$ and $f_{m/i}$ are the frequency and momentum basis functions. The basis functions are Fourier modes in both cases with the frequency basis sets modified to scale with the flow and cover the entirety of the imaginary time axis. Applying this expansion, the flow for the self-energy becomes
\begin{align}
\partial_\Lambda \Sigma^{\lambda}(s_{ph})=\textbf{g}(s_{ph})\textbf{L}_\Lambda^{ph}(s_{ph})\textbf{g}(s_{ph})
\end{align}
and the flow for the e-ph vertex is given by
\begin{align}
\partial_\Lambda \textbf{g}(s_{ph})=\textbf{g}(s_{ph})\textbf{L}_{\Lambda}^{ph}(s_{ph})\textbf{$\Delta$}(s_{ph})
\end{align}
where $L_\Lambda^{ph}$ represents an exchange propagator constructed from the electron propagator ($L[\partial_\Lambda(\mathcal{G}\mathcal{G})]$) and $\Delta$ is the two-particle vertex ($\Gamma$) expanded in the particle-hole channel. The above expressions are matrix multiplications with $\textbf{g}$ corresponding to a $[1\times N_\omega N_K]$ matrix for all singular particle-hole frequencies. The two-loop equations for the fermion vertices have been outlined in previous works\cite{eberlin2Loop,kugler2018multiloop}. The contributions to the e-ph vertex at higher loop orders can be accounted for by the derivative of the two-particle vertex. In particular, the projection of the fermion flows in the particle-particle ($\dot{\Phi}^{pp}$) and particle-hole-exchange ($\dot{\Phi}^{phe}$) channels contribute to the particle-hole vertex at the two loop level($\Delta^{2-L}=P(\dot{\Phi}^{pp})+P(\dot{\Phi}^{phe})$). This adds to the one-loop flow above as
\begin{align}
\partial_\Lambda \textbf{g}^{2-L}(s_{ph})=\textbf{g}(s_{ph})\textbf{L}_{F,\Lambda}^{ph}(s_{ph})\textbf{$\Delta$}^{2-L}(s_{ph})
\end{align}
with $L_{F,\Lambda}^{ph}$ corresponding to the full exchange propagator ($L[\mathcal{G}\mathcal{G}]$). Ultimately, the momentum structure of the e-ph coupling and the dispersion of the phonon introduce electron-electron interactions that drive charge fluctuations with a variety of momentum structures. To allow for deformations to the electron and phonon self-energies and an unbiased treatment of the charge fluctuations, the results throughout this work were constructed with the litim regulator implemented over the frequency axis of the electron propagator\cite{litim2001optimized}.

\begin{figure}
\centering
\includegraphics[scale=0.35]{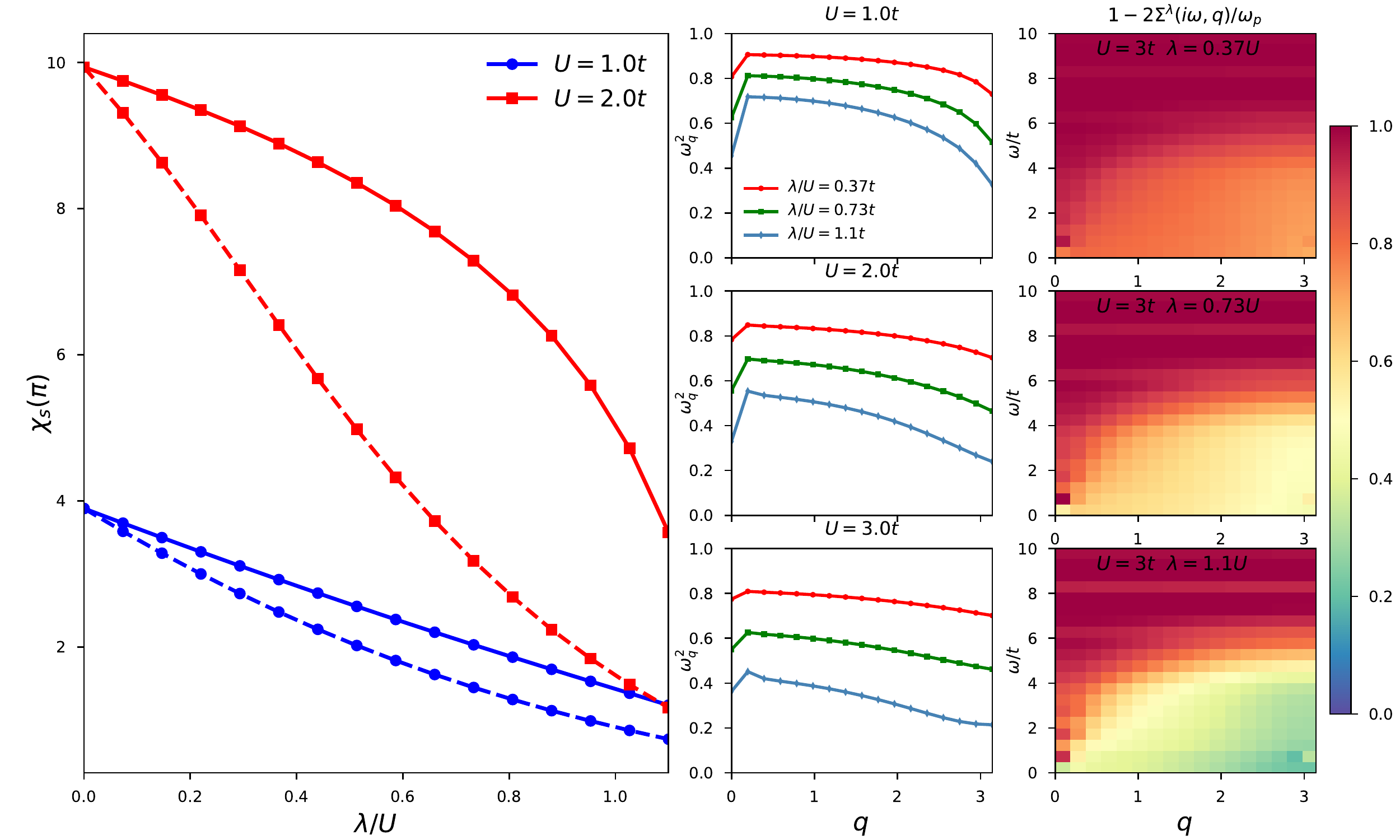}
\caption{The static anti-ferromagnetic susceptibility as a function of the electron-phonon coupling ($\lambda$) for different values of Hubbard coupling ($U$) in the HH model. Calculations were performed at 2-loop with $T=0.02t$ on a $32$-site lattice at a resolution of $N_\omega=4$,$N_k=4$ with the f-fRG(solid) and eph-fRG(dashed). The renormalized phonon dispersion and the self-energy of the phonon for different couplings calculated via the eph-fRG is shown on the right.}
\label{ephMethods}
\end{figure}

The main observables we use for studying the phases in the e-ph models are the spin, charge and superconducting correlators. From these one can construct the static susceptibility
and structure factors for ordering in the three channels. In parameter regimes with symmetry breaking instabilities, the flow has to be stopped due to a diverging interacting vertex, which limits us to correlators constructed from the partially integrated vertex at the critical scale $\Lambda$. In all cases we search for local orders with a profile characterized by the form factor $f_\mathcal{O}$. The spin and charge susceptibilities for a nesting vector $\vec{q}$ at a frequency $\Omega$ are given by
\begin{align}
\chi_{c/s}(\Omega,\vec{q})&=\sum_{p_1,p_2,\sigma_1,\sigma_2}\langle s_{\sigma_1}f_{\mathcal{O}}(p_1)c_{p_1,\sigma_1}^\dagger c_{p_1+p_q,\sigma_1}\times\nonumber\\
&s_{\sigma_2}f_{\mathcal{O}}(p_2)c_{p_2,\sigma_2}^\dagger c_{p_2-p_q,\sigma_2} \rangle_c
\end{align}
with the form factor corresponding to phase of interest. We investigate forms for all harmonics associated with the square lattice. Explicitly, searches for possible $BOW$ order along the x,y or z axes can be conducted with the factors $f_{\mathcal{O}}=\sin(p_i)$. The frequency content of the flow allows the calculation of the structure factor associated with a particular $\chi (S_\mathcal{O}=\frac{1}{\beta}\sum_\Omega\chi_\mathcal{O}(\Omega,\vec{q}))$ which can be ideal for determining phase boundaries. Preliminary transition lines constructed by direct comparison between susceptibilities can be supplemented (especially in cases with divergent flows) with studies of the structure factor. For cases where runs over different lattice sizes are possible, we extrapolate the transition lines for changes in the dominant susceptibility to the infinite lattice.

\begin{figure}
\centering
\includegraphics[scale=0.62]{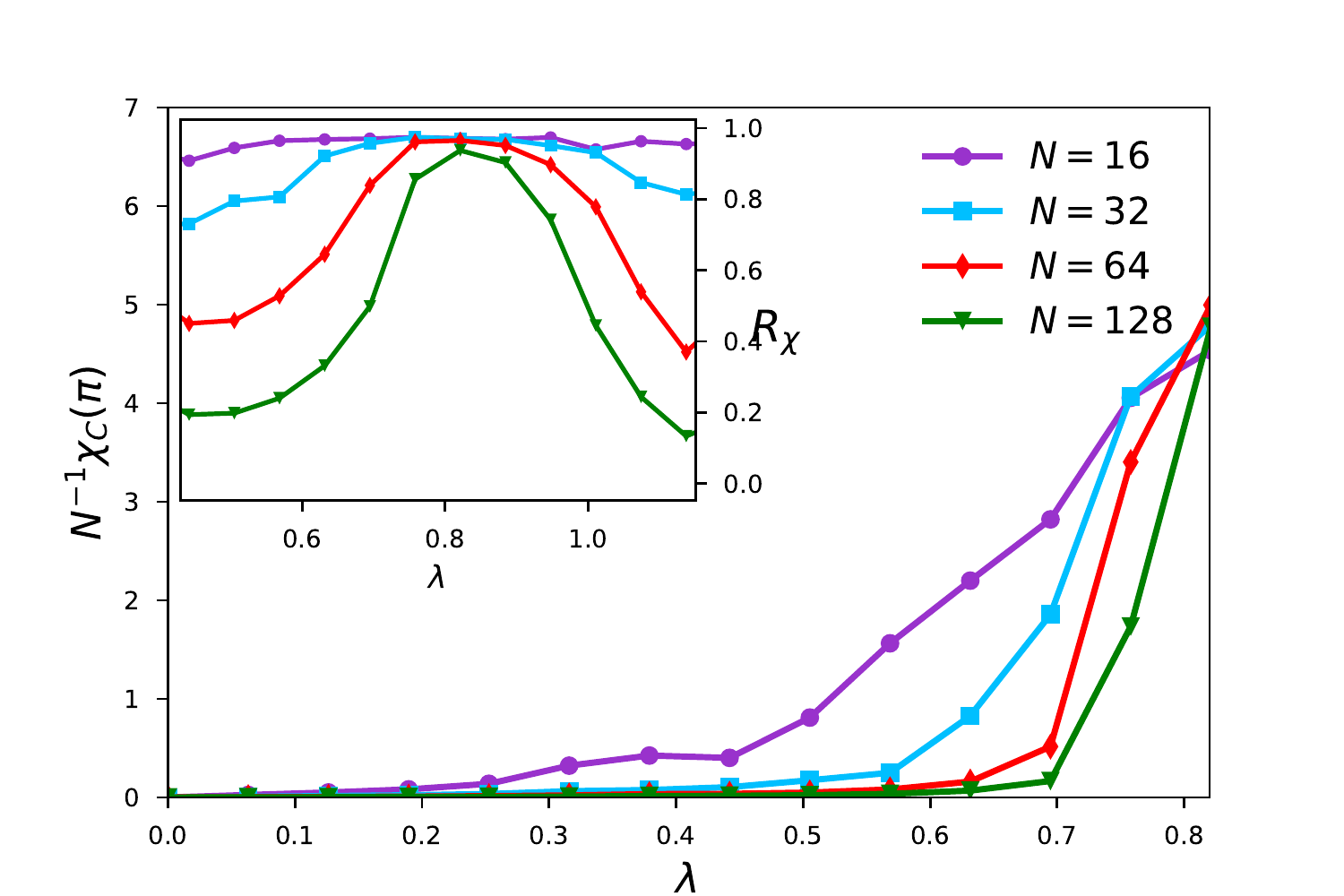}
\caption{The $CDW$ susceptibility of the Holstein model as a function of the e-ph for different lattice sizes at $T=0.02$ at a resolution of $N_\omega=4$,$N_k=5$. The inset shows the charge correlation ratio of the susceptibilities.}
\label{holstein1Dscaling}
\end{figure}

\section{The Hubbard-Holstein model}
\label{holsteinFRG}

\begin{figure}
\centering
\includegraphics[scale=0.57]{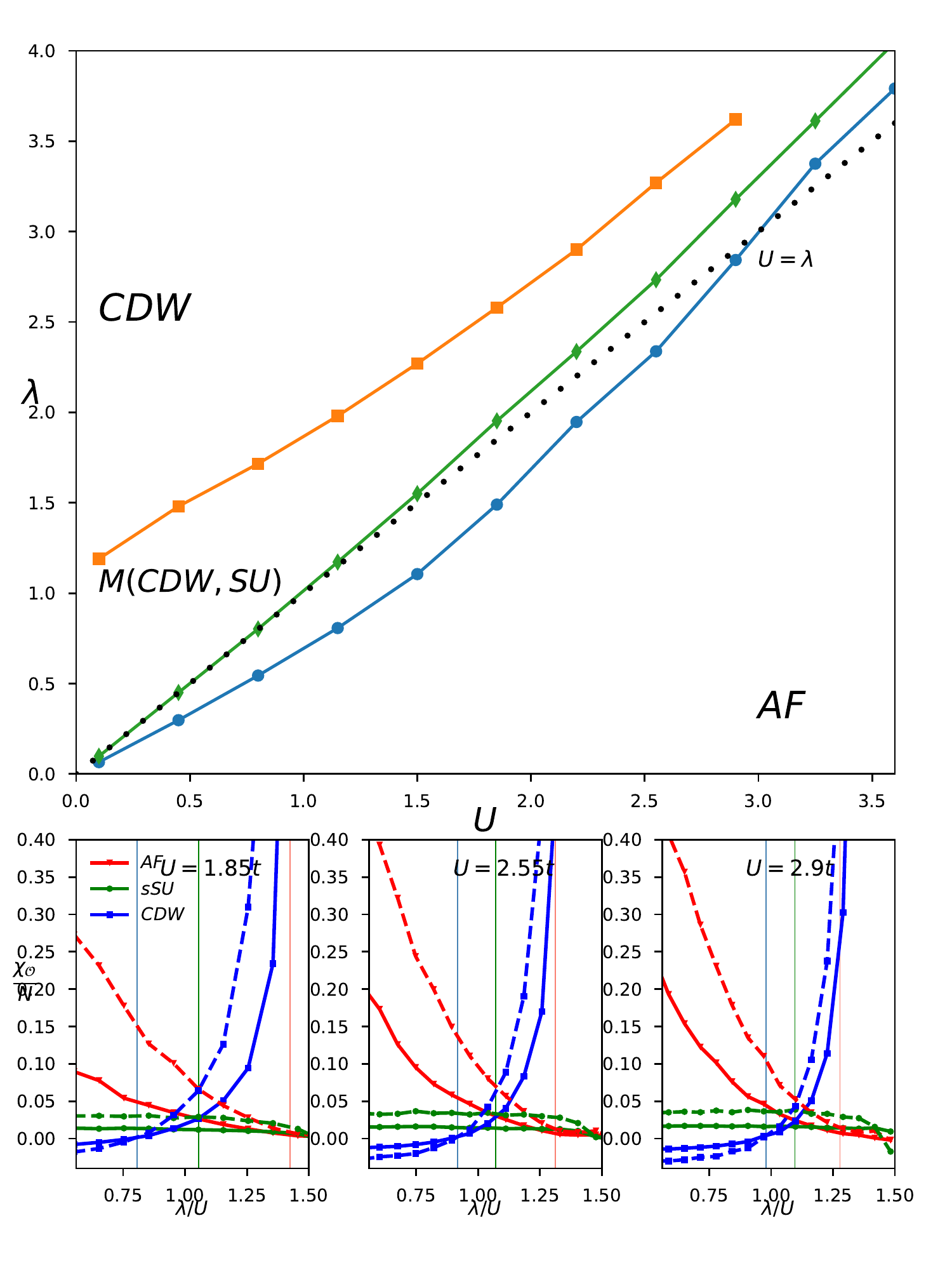}
\caption{The ground states of the HH model at half filling via the two-loop fRG with $\omega_0=t$. The transition from dominant charge to spin fluctuations is shown in green with metallic phase constructed by finite-size scaling of response in lattices of $N\leq 256$. The mean-field transition line ($U=\lambda$) is plotted for reference. The susceptibilities for $N=32$(dashed) and $N=64$ along with the transition lines is shown in the lower panel.}
\label{phaseDiagHH}
\end{figure}

The Hubbard-Holstein (HH) model is a prototype Hamiltonian for capturing the interplay between e-e and e-ph interactions. The model describes the coupling of fermions interacting via the Hubbard coupling to a non-dispersive optical phonon. Studies of the model show the expected antiferromagnetic phase ($AF$) for large e-e interactions ($U$) and a charge density wave ($CDW$) phase for strong e-ph coupling ($\lambda$) with a metal in the transition between the phases. In both one and two dimensions the metallic phases show strong superconducting correlations with 1D DMRG studies showing a metallic phase in the Holstein model ($U=0$) up to $\mathcal{O}(1)$ values of e-ph coupling for various values of the phonon frequency\cite{tezuka2007phase,jeckelmann1999metal}. Monte Carlo studies of the model in 2D show a $CDW$ phase at $U=0$ for any values of the e-ph coupling\cite{costa2020phase,wang2020zero,ohgoe2017competition}. These studies find that the metallic phase still exists in 2D, albeit with a reduced domain and weaker sensitivity to the frequency of the Holstein phonon. In terms of a full description of the model, the studies in 2D have yet to address the impact of doping and of the phonon frequency on the metallic phase. In 1D, DMRG studies of the model find a significant enlargement of the metallic phase as the phonon frequency is increased. Additionally, the 1D study finds the $CDW$ phase persists even as doping destroys the nested Fermi surface, and large levels of doping are required ($x>0.1$) before the superconducting fluctuations dominate the density wave. In what follows we will study the role of these parameters in the two dimensional models. These models offer a rich playground that captures the interplay between superconducting and incommensurate spin fluctuations seen in the Hubbard model with the charge and superconducting response created by the retarded interaction with the phonon mode.

\begin{figure}
\centering
\includegraphics[scale=0.41]{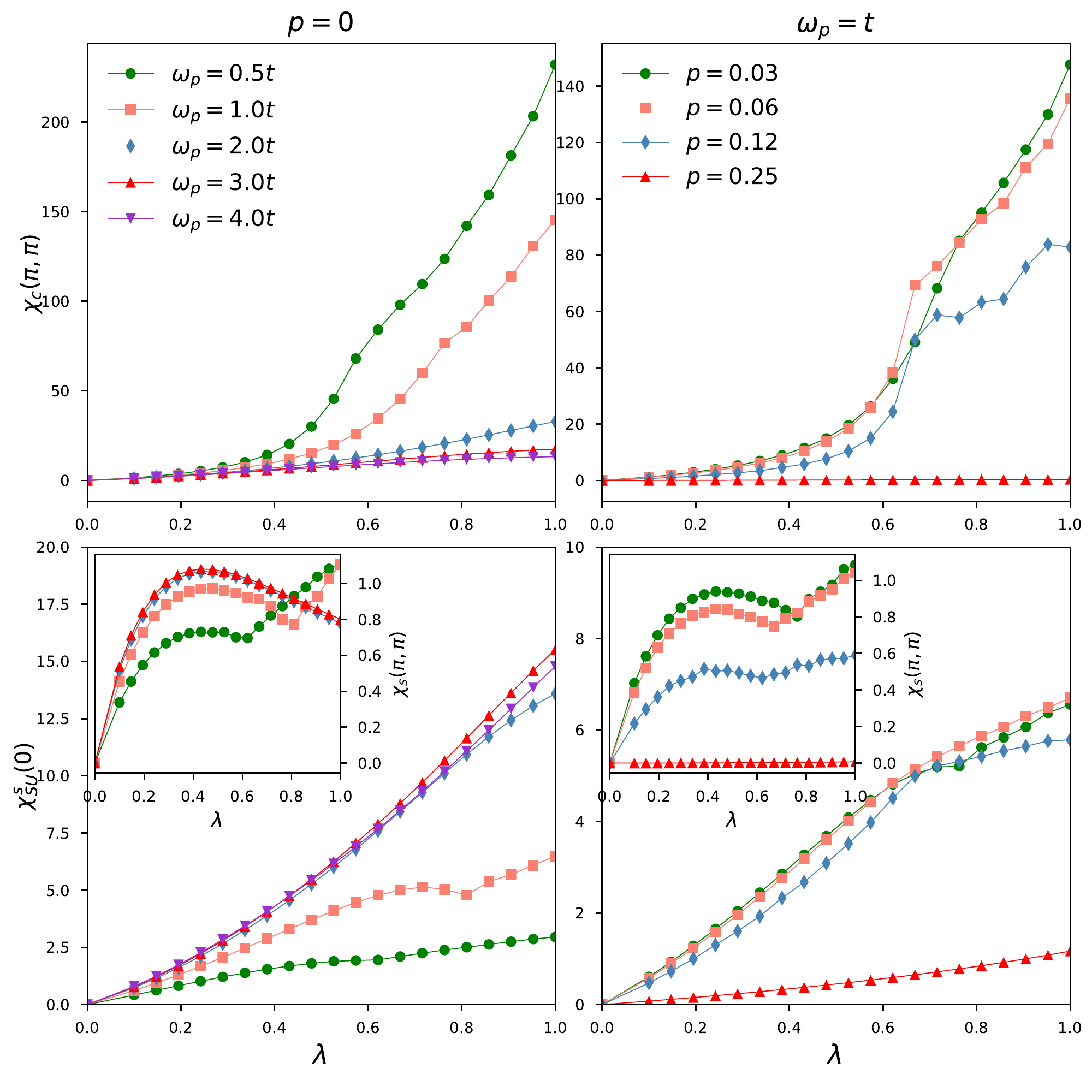}
\caption{The charge (top), spin (inset) and s-superconducting (bottom) susceptibilities of the 2D Holstein model at various phonon frequencies (left) and doping levels (right) on a $16\times 16$ lattice at $\beta=32$.}
\label{holsteinWX}
\end{figure}
\begin{figure*}
\centering
\includegraphics[scale=0.45]{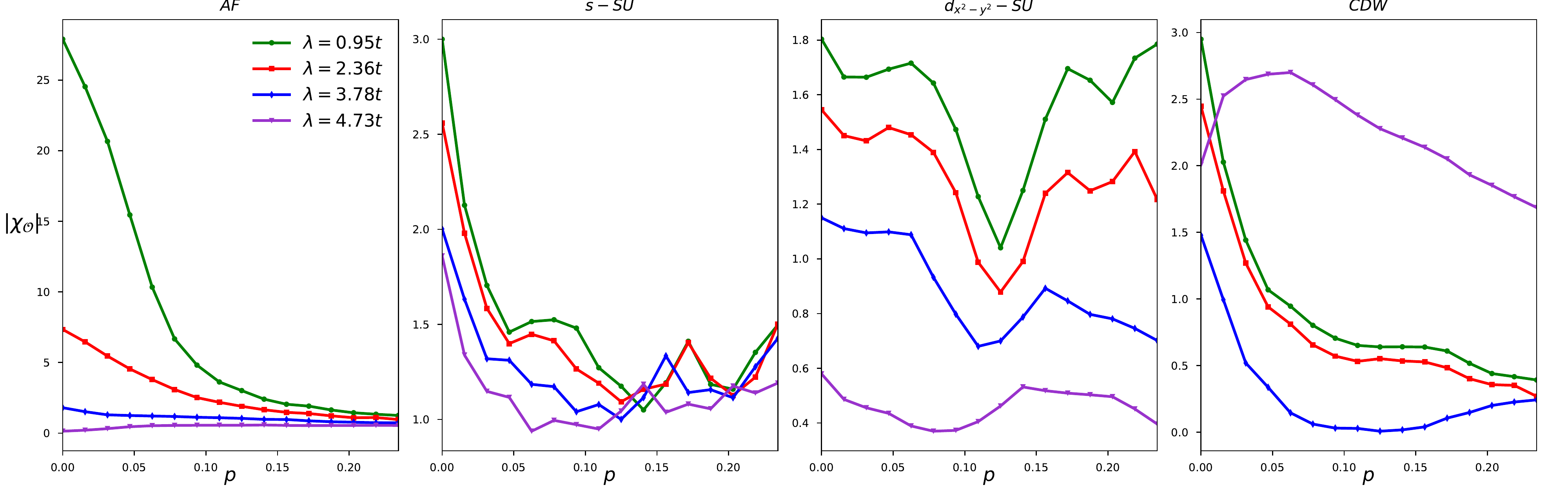}
\caption{The antiferromagnetic, superconducting ($s,d_{x^2-y^2}$) and charge susceptibilities of the 2D HH model as a function of doping for different values of the e-ph coupling at $U=4t$. Calculations were performed at the 2-loop level with $T=0.02t$ on a 16$\times$16 lattice at a resolution of $N_\omega=4$,$N_k=3$. }
\label{dopingSus2DHH}
\end{figure*}

The presence of the metallic phase with charge and superconducting correlations complicates the usual instability analysis carried out for vertices constructed by the fRG.
Quantum Monte Carlo studies of the one dimensional Holstein model found that corrections for finite-size systems compounded with the exponentially small gap make the determination of the metallic domain difficult to determine numerically\cite{tam2011validity}. With this in mind, we separate charge fluctuations in the metallic phase from the $CDW$ phase by performing a finite size scaling analysis on the charge correlation ratio\cite{binder1981critical}. This approach enables the determination of the $CDW$ phase boundary with the charge susceptibilities for various lattice sizes as input. Though it offers numerous computational advantages, our current limitation to a $20\times 20$ lattice in two dimensions requires an alternative solution. DMRG studies of the one dimensional model find charge and superconducting correlations in the metallic phase decaying with power law behavior, whereas in the $CDW$ phase charge fluctuations show little decay with other correlations suppressed exponentially. So we will supplement the susceptibility of the various orders with the decay behavior of the correlators to construct a phase diagram of the models. Given the previous studies of the model we will focus on instabilities of $s$, $ext-s$ and $d_{x^2-y^2}$ type in the three channels.

The dispersion relation for electrons in the HH model is determined by the nearest neighbor hopping, $\xi_k^e=-2t(\cos(k_x)+\cos(k_y)+...)$, with the system coupled to a non-dispersive phonon mode ($\Omega^\lambda=\omega_0$). Similarly, the e-ph interaction is local, $g_{k,q}=g_0$, and couples equally to all momentum modes.  We begin with a study of the 1D system in order to evaluate the benefit of retaining the flow of phonon vertices. Beyond enabling access to the phonon self-energy, flowing the e-ph vertex modifies the electronic vertex as the effective e-e interaction changes with the flow. A comparison between the spin susceptibility calculated by the two approaches as we approach the $CDW$ transition is shown in Fig.\ref{ephMethods}. Suppression of spin fluctuations is expected as we approach the transition, as both the metallic and $CDW$ phases show little spin response. The flow of the e-ph vertex leads to a stronger suppression of the spin response with the phonon self-energy showing large deviations of $\mathcal{O}(t)$ for an initial phonon frequency, $\Omega=t$. The changes to the phonon self-energy occur over a wide frequency window $\sim\mathcal{O}(W)$ centered at the $\pi$-phonon corresponding to the $CDW$ phase. From it we can extract the  renormalized phonon dispersion, $\omega_{p}^2=\omega_{p,0}^2(1-2\Sigma_k^\lambda/\omega_{p,0})$, which shows the softening of the phonon modes as we approach the transition. For the pure Holstein model ($U=0$) both Monte Carlo and DMRG studies find charge order setting in at $\lambda=t$ for a phonon mode with dispersion $\Omega=t$ \cite{hardikar2007phase,tezuka2007phase}. The fRG results for the charge response of the Holstein model is shown in Fig.\ref{holstein1Dscaling}. Analysis of the charge susceptibility and correlation ratio ($R_{\chi(q)}=1-\chi_{q+\delta q}/\chi_q$) find the transition at $\lambda=0.81t$\cite{binder1981critical}. Given the various truncations of the fRG hierarchy and two-particle vertex and given that we are trying to separate two phases showing a strong charge response, discrepancies can be expected. As the discrepancy is reduced as we move to larger system sizes and increase the momentum and frequency resolution of the vertex, the truncations appear to be the likely source of the disagreement. Further, our current calculations are limited to the two-loop level and calculations at higher loop levels may be needed to enable a better resolution of the transition.

\begin{figure}
\centering
\includegraphics[scale=0.4]{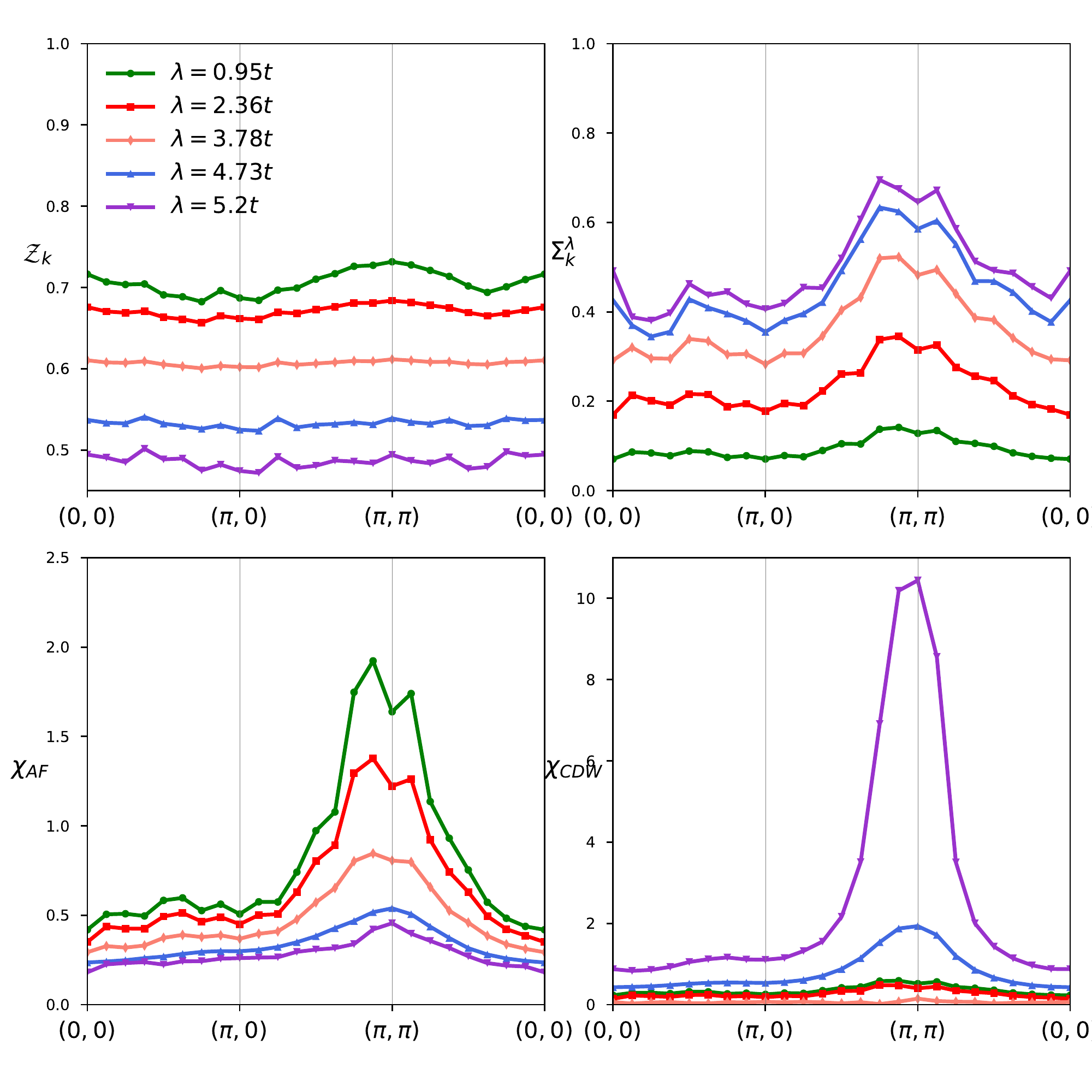}
\caption{The quasiparticle weight ($\mathcal{Z}_k$), the phonon self-energy ($\Sigma_q^\lambda$) spin and charge susceptibilities as a function of the e-ph coupling at $U=4t$ of the doped ($p=0.18$) 2D HH model on a 16$\times$16 lattice with $T=0.02t$. }
\label{susSEXL}
\end{figure}

Constructing the phase diagram of the HH model requires separating the metallic, charge-ordered and antiferromagnetic regions. We identify the start of charge correlations in the the $U-\lambda$ plane by utilizing the correlation ratio to find the intersection between charge responses for different lattices. A similar procedure can be performed on the spin response to determine the end of the antiferromagnetic regime. The intervening metallic region lacks long-range charge order which should lead to a charge response that decreases with system size. As the system transitions into the $CDW$ phase the charge response saturates, leading to a divergent susceptibility. We use these points of intersection to construct the phase diagram shown in Fig.\ref{phaseDiagHH}. Our results show a qualitative agreement with previous RG and DMRG studies with much of the error appearing in the transition from the metallic phase to the $CDW$ phase\cite{bakrim2015nature,tezuka2007phase}. Given that the fRG flow leads to a divergent charge response, this discrepancy is expected. The exponential suppression of spin fluctuations in the $CDW$ phase allows us to use the termination point for the spin response as an estimate to the end point of the metallic phase. The results for the other transition from an $AF$ to a metallic phase show much better fidelity, which is consistent with previous fRG studies, as the spin gap is closed in one dimension and spin fluctuations show power law scaling.

\begin{figure}
\centering
\includegraphics[scale=0.43]{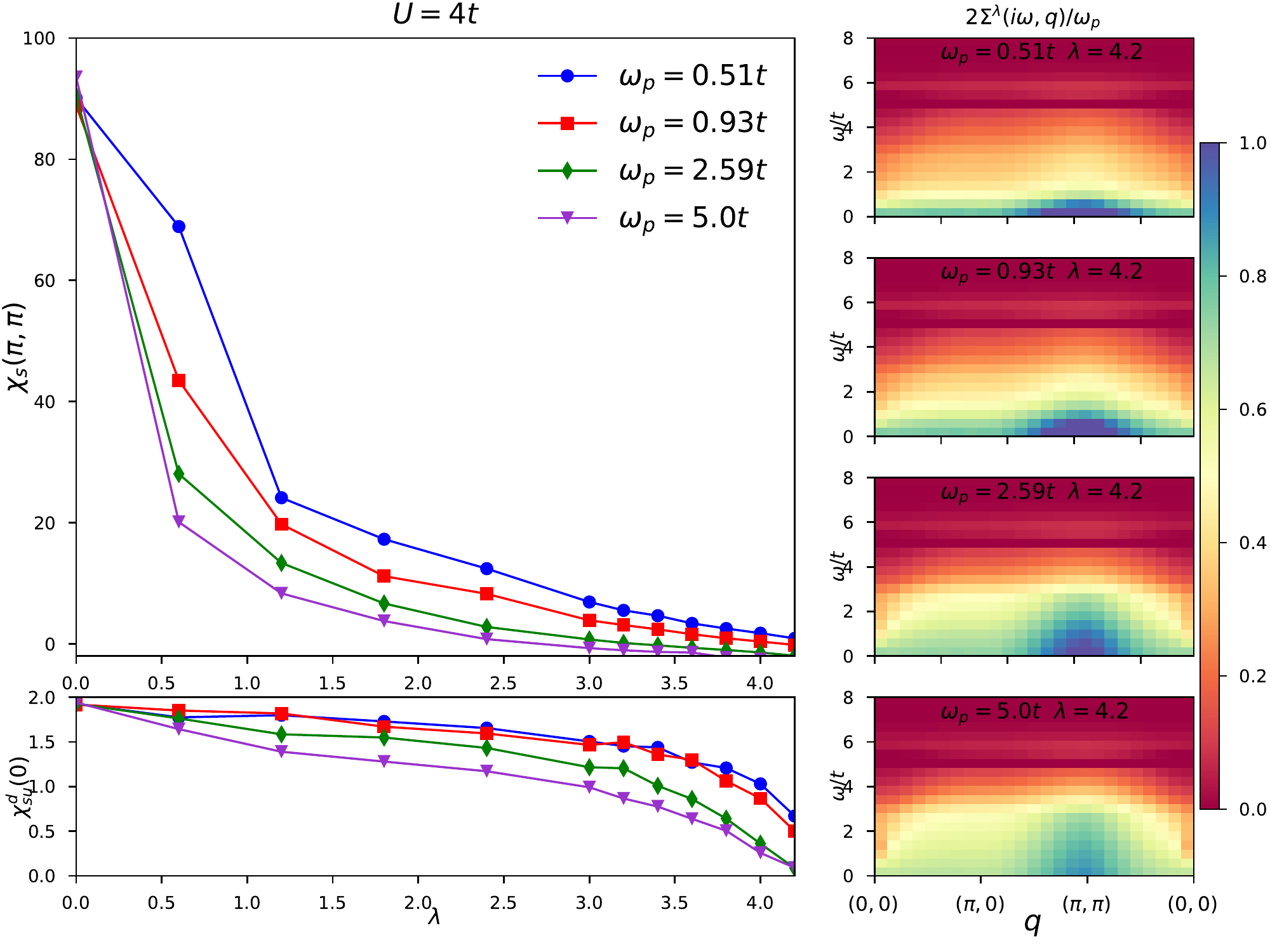}
\caption{The static $(\pi,\pi)$-spin, $d_{x^2-y^2}-SU$ and $s-SU$ susceptibilities of the 2D HH model as a function of the e-ph coupling for different phonon frequencies at $U=4t$ for $N_\omega=4$,$N_k=2$ on a 16$\times$16 lattice. The phonon self-energy for the different frequencies is shown on the right ($\lambda=4.16t$).}
\label{spinSus2DHH}
\end{figure}

\begin{figure}
\centering
\includegraphics[scale=0.57]{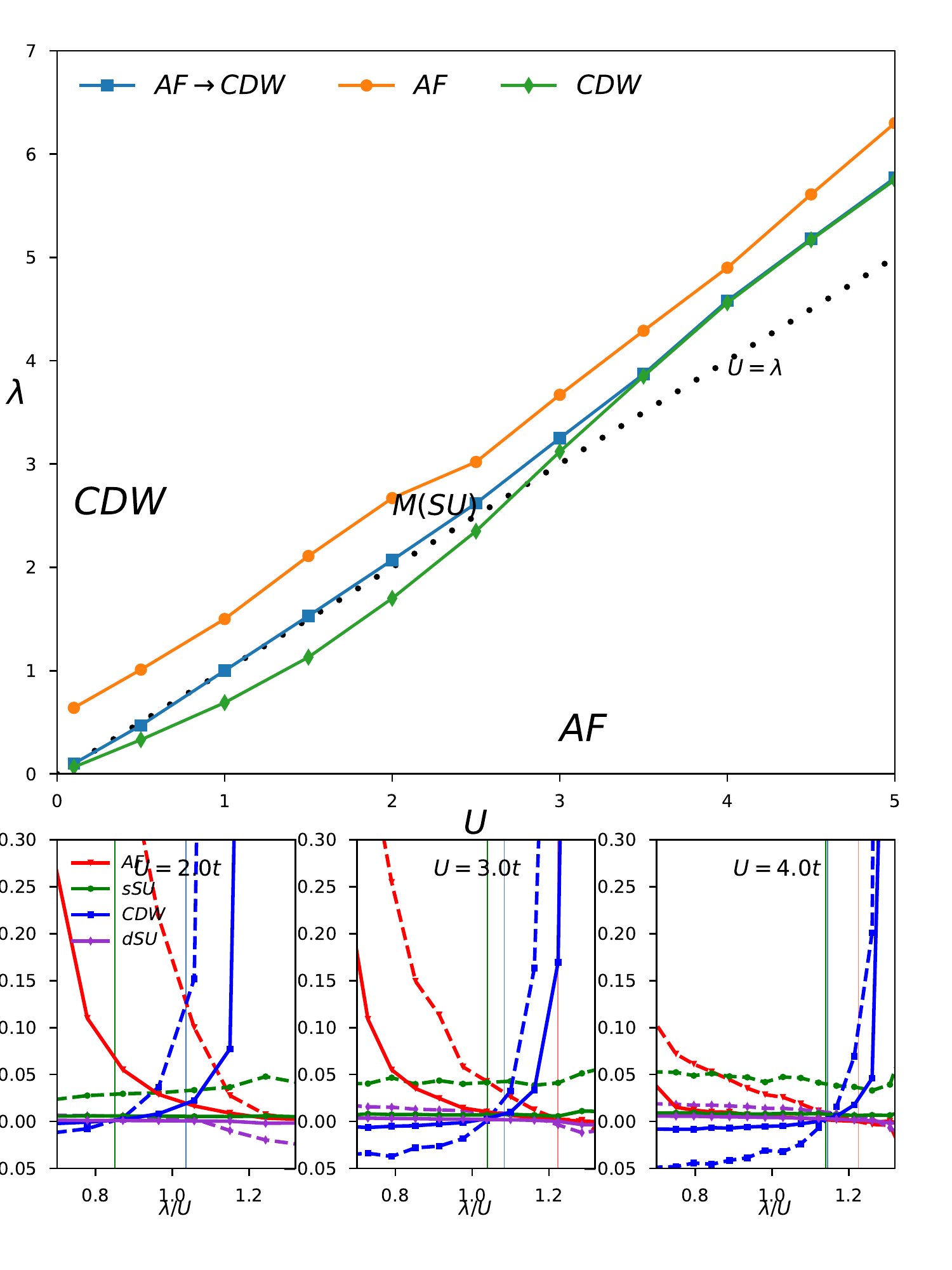}
\caption{The phase diagram of the 2D HH model at half filling via the two-loop fRG with $\omega_0=t$. The transition between the $AF$, $CDW$ and $SU$ orders are shown for the $16\times16$ lattice at $N_\omega=4$, $N_k=2$ and $T=0.02t$. The susceptibilities scaled with $N$ are shown in the lower panel as a function of $\lambda$ for an $8\times 8$ (dashed)and $16\times 16$ lattice. }
\label{phaseDiagHH2D}
\end{figure}

\begin{figure*}
\hspace*{-1cm}\includegraphics[scale=0.5]{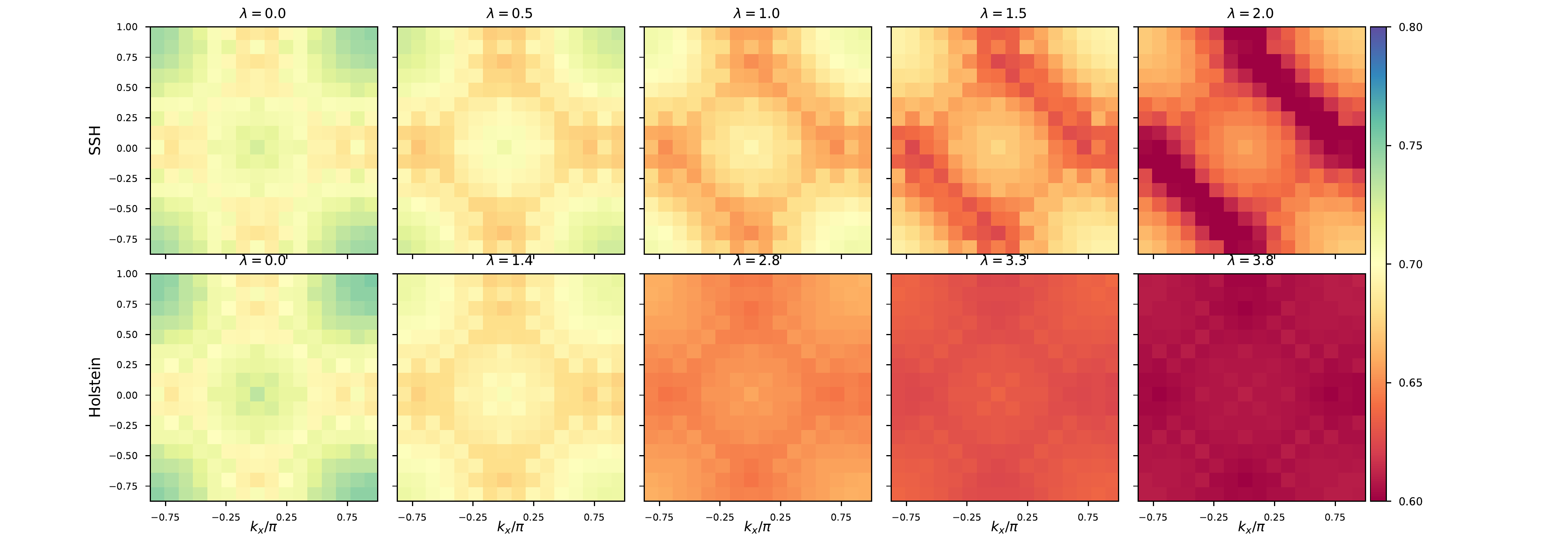}
\caption{The quasiparticle weight ($\mathcal{Z}_k=1-\partial_{i\omega=0}\Sigma(i\omega,k)$) of the 2D HH and PH models as a function of the e-ph coupling ($\lambda$) for a doped ($p=0.18$) system at $U=4t$. The phonon frequency ($\omega_0$) is set to the hopping ($t$) with the transition to $BOW$ occurring at $\lambda\approx t$. }
\label{qpSE}
\end{figure*}

In 2D, Monte Carlo studies of the model at half-filling find a shrunken metallic phase nestled between the $AF$ and $CDW$\cite{ohgoe2017competition,costa2020phase}. Given the need to account for the frequency modes of the vertex, the fRG at the two-loop level is currently limited to lattice sizes of $20\times 20$. With this in mind, we carried out studies of phases in the model with two lattices of linear dimensions $8$ and $16$. As we wish to study the self-energy of the phonon, we allow the e-ph vertices to flow independently. Important directions that have not been explored fully in previous studies of the model are the sensitivity of the metallic phase to doping and phonon frequency. Earlier variational Monte Carlo studies of the model find a strong response to phonon frequency with results at $\omega_p=8t$ showing a larger metallic phase\cite{ohgoe2017competition}. The decoupled fRG is versatile with respect to both parameters, so we can construct the susceptibilities of the system at a large range of doping levels and phonon frequencies. The performance of the fRG for different phonon frequencies can be further improved by adjusting the patching of the time domain we average over to decouple the frequency dependence of the vertex to the frequency of the phonon. Control over these two parameters allows us to explore the proposed lack of metallic phase in the Holstein model. Previous unbiased Monte Carlo studies of the model at low phonon frequencies\cite{costa2020phase,hohenadler2019dominant,weber2018two} ($\omega_p<t$) indicate the nonexistence of the phase, with numerical results restricting possible transition to $\lambda<0.61t$. To address this, we calculated the response of the 2D Holstein model at various doping values for different phonon frequencies. The results are shown in Fig.\ref{holsteinWX}. At low phonon frequencies ($\omega_p=0.5t,t$) the flow diverges at finite $\lambda$ with the strong charge response indicating a transition to a charge ordered phase. The flow is convergent for larger values of the phonon frequency with the charge susceptibility decreasing and approaching an increasing s-wave superconducting response. We extrapolated the charge response at various doping values and phonon frequencies to find the transition to the charge ordered phase for $\omega_p=t$ at $\lambda=0.64t$. For higher phonon frequencies ($\omega_p\geq 2$) the transition is beyond the parameter region ($0<\lambda<t$) considered indicating an expanding metallic phase.

\begin{figure}
\centering
\includegraphics[scale=0.59]{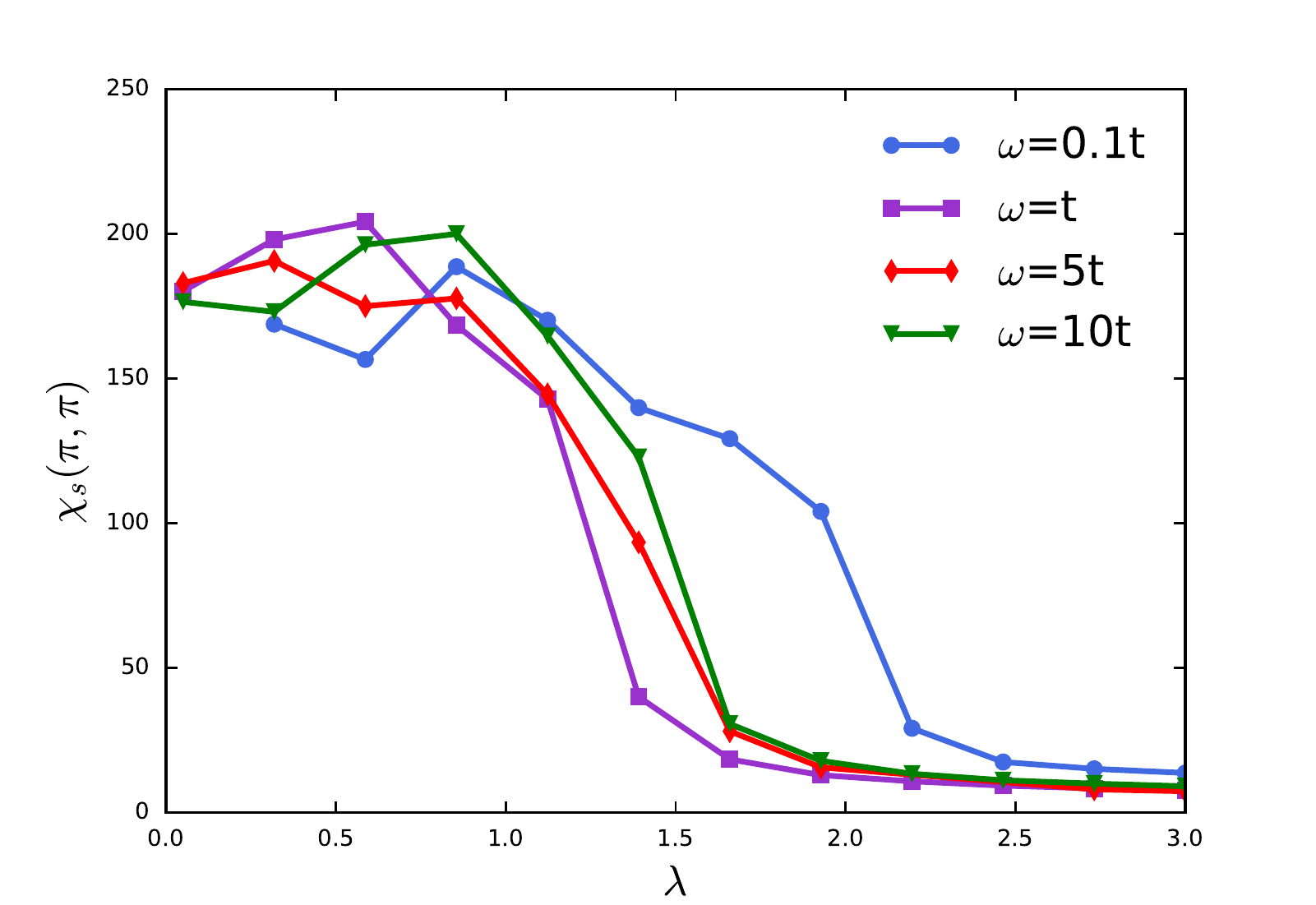}
\caption{The static $(\pi,\pi)$-spin susceptibility of the PH model as a function of the electron-phonon coupling for different phonon frequencies at $U=t$. Calculations were performed at 2-loop with $T=0.02t$ on a 16$\times$16 lattice at a resolution of $N_\omega=4$,$N_K=2$. }
\label{spinSusSSH}
\end{figure}

Doping the model at finite $U$ suppresses the strong spin correlations seen in the Hubbard model, allowing the spin-facilitated superconducting correlations to come to the forefront. The presence of superconducting fluctuations in the metallic phase of Holstein model that expands with phonon frequency leads to the expectation of superconducting order as the combined system is doped. The response of the system to doping is shown in Fig.\ref{dopingSus2DHH}. As expected the anti-ferromagnetic correlations in the system are suppressed by increasing the coupling to the phonon mode and the destruction of the perfectly nested Fermi surface due to the doping. The $CDW$ order generated at large values of $\lambda$ shows little sensitivity to doping, in line with the DMRG results for the 1D system. The s-type superconducting response is more telling, as unlike the d-type order which shows little change as the system is doped away from half filling, the s-type is significantly reduced as a function of doping. This suggests that the superconducting fluctuations seen in the metallic phase are formed due to the interplay between the e-ph coupling and nesting of the Fermi surface, so that a metallic phase is the likely to dominate the regime of low doping. Despite the reduction of the spin fluctuations due to the e-ph coupling, the d-type superconducting response remains large at  $p>0.15$ suggesting a transition from $d-SU$ to a metal to the $CDW$ in this doping regime. Incommensurate correlations are the rule in the doped regime with both the $AF$ susceptibility and the phonon self-energy showing peaks at incommensurate wave vectors. A large enough e-ph coupling suppresses these incommensurate fluctuations, with the $CDW$ order forming at large $\lambda$, as shown in Fig.\ref{susSEXL}. Large values of the e-ph coupling lead to a stronger renormalization of the quasiparticle weight ($\mathcal{Z}_k$) but the interaction destroys the nodal structure seen in the quasiparticle weight of the Hubbard model. This is in line with charge ordering driven by the Holstein e-ph interaction which couples equally to all momentum modes.

The impact of the phonon frequency on the system response is shown in Fig.\ref{spinSus2DHH} for the 2D HH model. Increasing the frequency of the phonon suppresses spin fluctuations with physics akin to the anti-adiabatic limit ($\omega_0\rightarrow\infty$) leading to a sharper transition out of the $AF$ phase. In the large frequency limit, the electronic interactions have little effect on the phonon modes which quickly suppress the spin response allowing the superconducting fluctuations to come to the forefront. In this limit the system can be approximated by a Hubbard model with an interaction reduced by the e-ph coupling, which is in accordance with the observed weak spin response\cite{ohgoe2017competition}. The evolution of the system from the anti-adiabatic limit can be clearly seen in the phonon self-energy. In the $\omega_p\sim W$ range we see a weak softening of the $(\pi,\pi)$-mode associated with charge ordering. As the frequency of the phonon is lowered the response broadens with the self-energy at $\omega_p=0.5t$ showing a response at all momentum modes and the system retaining spin fluctuations into the charge-ordered phase.

\begin{figure}
\centering
\includegraphics[scale=0.4]{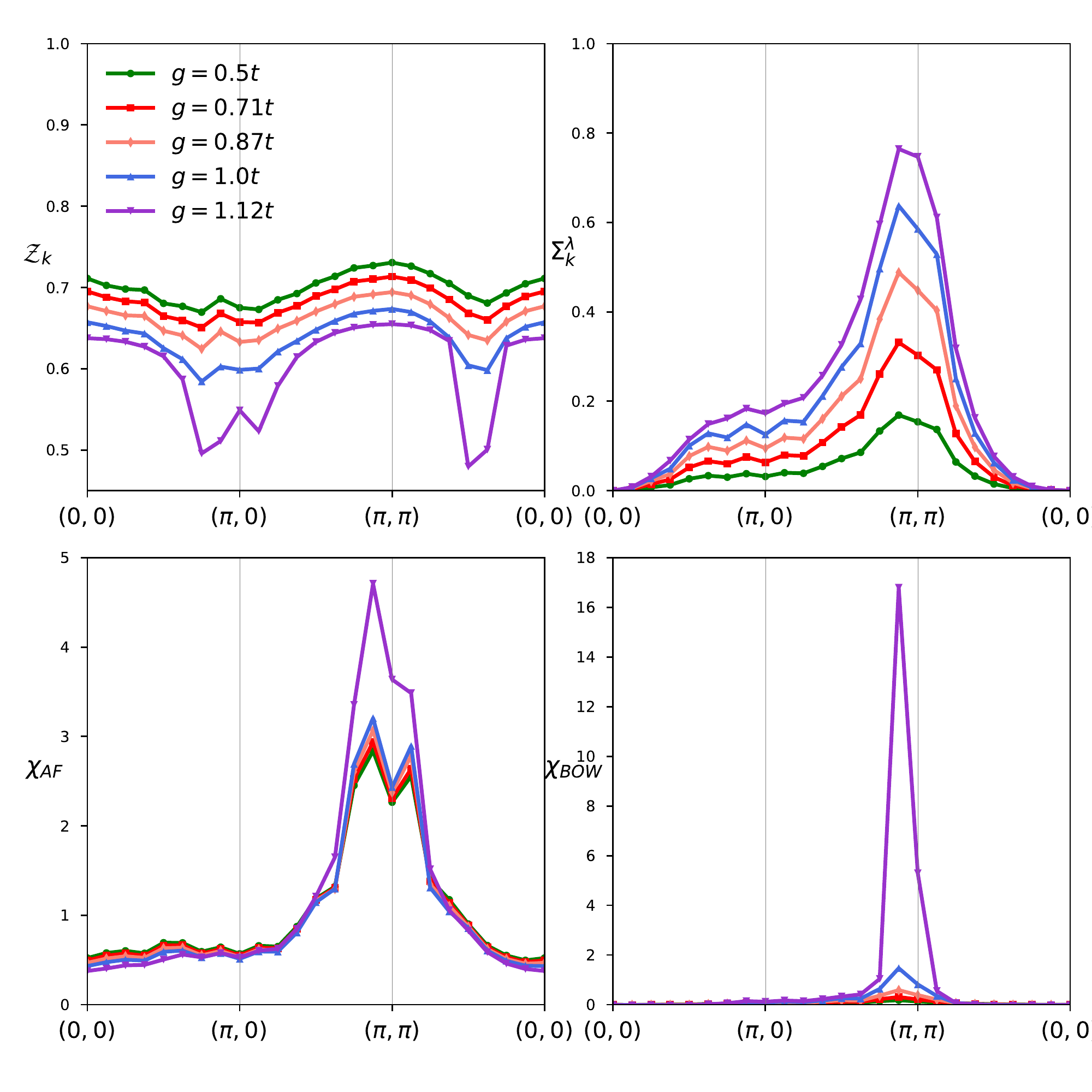}
\caption{The quasiparticle weight ($\mathcal{Z}_k$), the phonon self-energy ($\Sigma_q^\lambda$) spin and charge susceptibilities as a function of the e-ph coupling at $U=4t$ of the doped ($p=0.18$) 2D PH model on a 16$\times$16 lattice with $T=0.02t$, $N_{\omega}=6$ and $N_k=2$.}
\label{dopSusSSH}
\end{figure}

The phase diagram of the 2D HH model constructed in accordance with its 1D counterpart is shown in Fig.\ref{phaseDiagHH2D}. The stronger response in the spin channel leads to a larger AF phase and a further shift away from the mean field $U=\lambda$ line. A background of $s-SU$ fluctuations is present for much of the $U-\lambda$ domain but remains sub-dominant to the $CDW$ response as the e-ph coupling suppresses $AF$ correlations. Both in the $8\times 8$ and $16\times 16$ lattices the $s-SU$ susceptibility does not increase with the coupling which confirms the expected metallic phase populated by superconducting fluctuations. Though higher phonon frequencies lead to a stronger superconducting response for the frequencies considered ($\omega_p<5t$) we find and even stronger $CDW$ response. At half filling the $d-SU$ is suppressed by the e-ph coupling, which is in line with a suppression of the $AF$ which serves as its primary driver.

\section{The extended Hubbard-Peierls Model}
\label{sshFRG}

\begin{figure}
\centering
\includegraphics[scale=0.4]{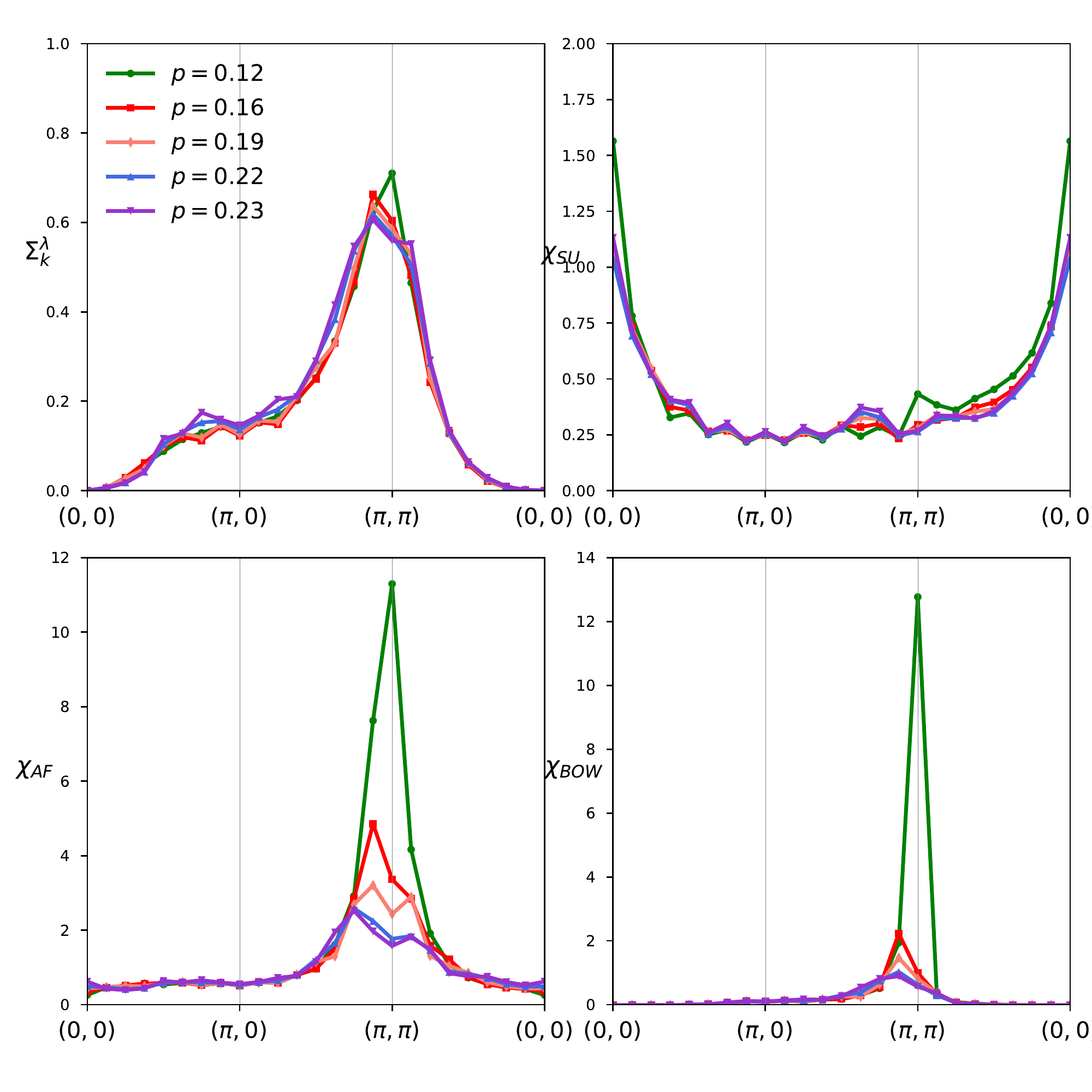}
\caption{The phonon self-energy ($\Sigma_q^\lambda$) ,spin and charge susceptibilities as a function of the doping at $U=4t$, $g=t$ of the 2D PH model on a 16$\times$16 lattice with $T=0.02t$, $N_{\omega}=6$ and $N_k=2$.}
\label{dopPSusSSH}
\end{figure}

\begin{figure*}
\hspace*{-1.5cm}\includegraphics[scale=0.63]{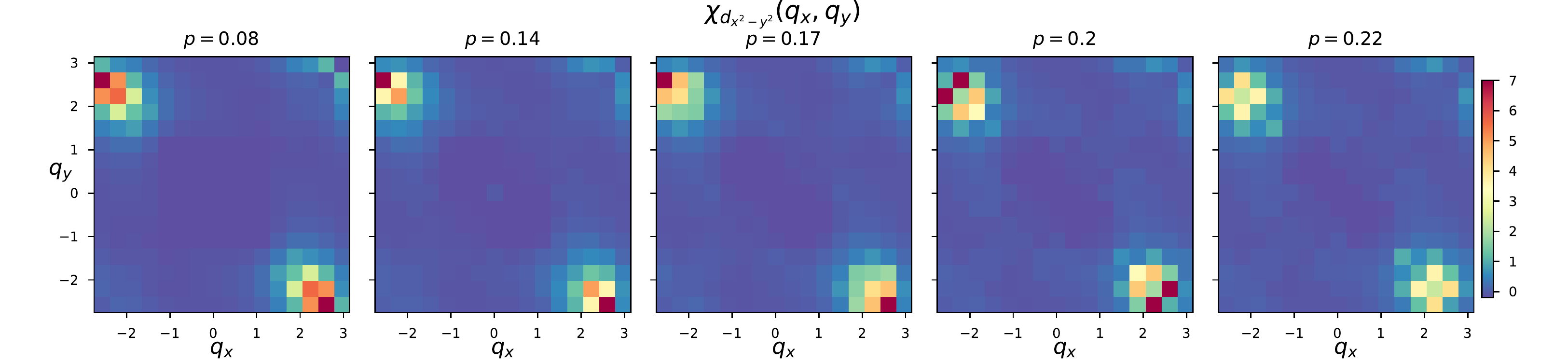}
\caption{The $d_{x^2-y^2}$-Charge susceptibility of the 2D PH model for $g=t$ at $U=4t$ at various values of doping on a 16$\times$16 lattice with $T=0.02t$, $N_{\omega}=6$ and $N_k=2$ }
\label{dChargeSSH}
\end{figure*}

\begin{figure}
\centering
\includegraphics[scale=0.28]{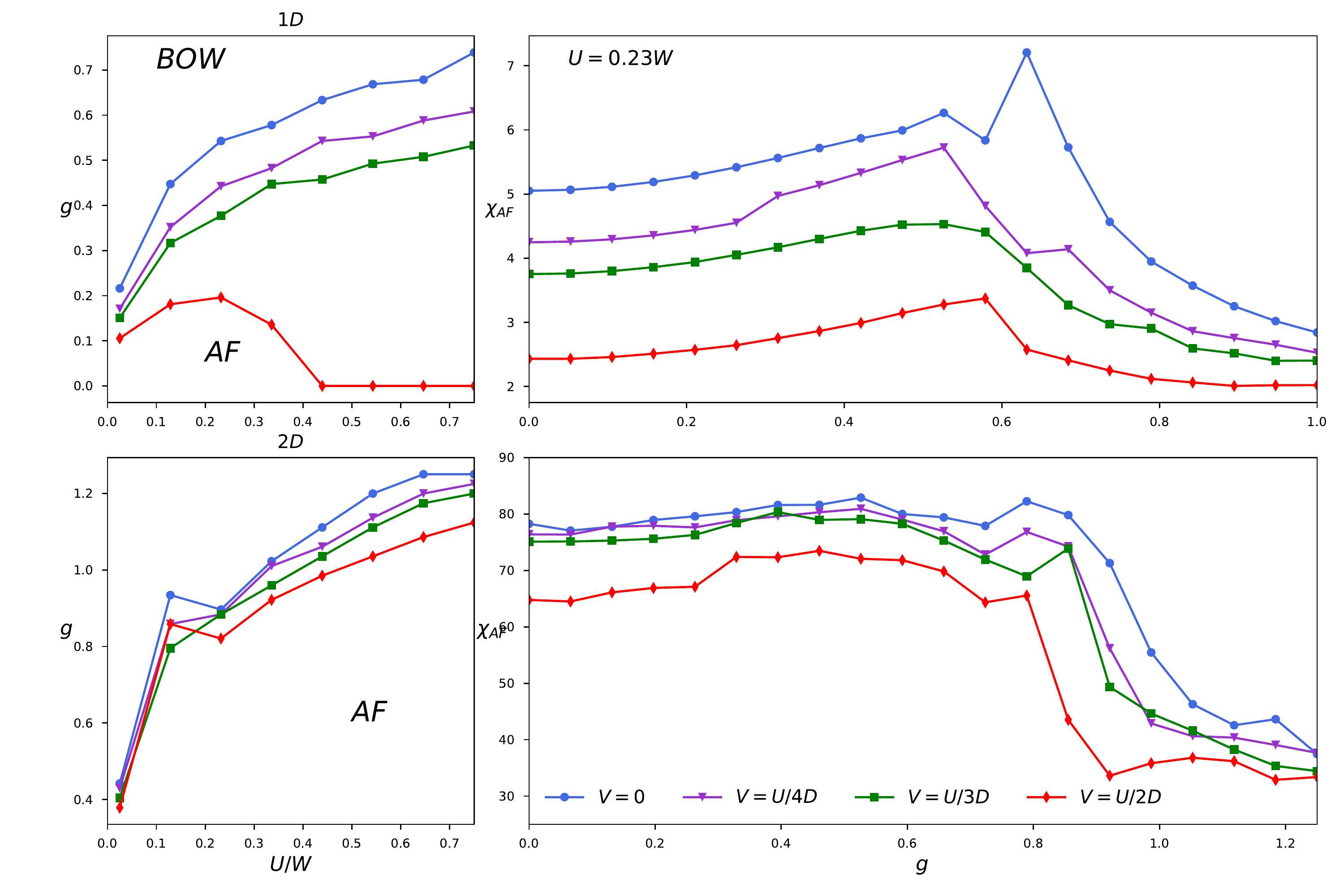}
\caption{The phase diagram of the Extended PH model as a function of extended Hubbard coupling ($V$) for SSH phonon with $\omega=t$ in one and two dimensions calculated with $N_\omega=6$ and $N_k=2$ at half filling. The spin response of the system across the $AF\rightarrow BOW$ transition is shown on the right.}
\label{phaseDiagSSH}
\end{figure}

Distortions of an elastic lattice due to coupling to electronic modes drive the physics in a variety of quasi one-dimensional materials, including organic charge-transfer solids and perovskite systems. The study of this interplay between electrons and lattice vibrations is typically modeled by the Su-Schrieffer-Heeger (SSH) Hamiltonian\cite{su1979solitons}. The Peierls instability in the SSH model shows a dimerization of the lattice with a bond-ordered charge density wave ($BOW$) at arbitrary values of the coupling for 1D
systems.  Quantum Monte Carlo studies of the model in two dimensions find a similarly dimerized lattice with the charge ordering on the bond along the x or y axis, albeit with the transition to the $BOW$ phase occurring at a finite value of the e-ph coupling ($g_c\approx0.67$)\cite{xing2021quantum}. Although the initial intent of the SSH Hamiltonian was as a model of dimerization in polyacetylene, a description of the material requires the inclusion of interactions between the electrons, as a large portion of the charge gap is due to these interactions\cite{jeckelmann1998mott}. Studies of an expanded SSH model with electronic interactions in 1D show a transition between the Peierls $BOW$ phase and an antiferromagnet as a function of interaction and phonon frequency\cite{sengupta2003peierls}. Renormalization group studies of the SSH model in one dimension capture this transition with the impact of the phonon frequency studied by a frequency dependent RG\cite{bakrim2015nature,bakrim2007quantum,caron1984two}. Recent studies of electronic interactions in the 2D SSH system find a similar transition from an $AF$ to a $BOW$ as a function of the e-ph coupling \cite{feng2021phase}. The former reference looked at extended Hubbard interactions expected in polyacetylene and found a transition between the phases showing robust competition, while the latter reference considered only a local Hubbard coupling and captured a transition showing weak dependence on the interaction.

The electronic sector of this model remains the same, but the elastic lattice leads to the phonons dispersing as $\xi_k^{ph}=\omega_0\sqrt{\sin(k_x/2)^2+\sin(k_y/2)^2}$. The phonon mode couples to the hopping of the electrons, leading to an e-ph interaction given by $g_{k,q}=ig(\sin(\frac{q_x}{2})\cos(k+\frac{q_x}{2})+(x\leftrightarrow y))$. The dispersive phonon mode and the momentum structure in the coupling lead to an extended effective interaction between the electrons. The strength of the effective e-e interaction can be defined by $\lambda_{SSH}=2g^2/\omega_0$ though interactions felt by the electrons vary with momenta. The extended Hubbard interactions in the model are accounted for by the term $V=2V(\cos(q_x)+\cos(q_y))$.

At half filling, coupling the Hubbard model to the SSH phonon leads to a response in the spin channel. Fig.\ref{spinSusSSH} shows the antiferromagnetic susceptibility for the model as a function of the strength of the e-ph coupling. We see that the coupling to the phonon mode initially enhances spin correlations for all values of the phonon frequency but once the system transitions to the $BOW$ phase, spin fluctuations show little enhancement. Models of materials normally occupy this low frequency parameter range ($\omega<<t$) with slow lattice vibrations and an e-ph coupling directly related to variations in the hopping integral due to lattice fluctuations. In this regime the coupling to the phonon mode can be expected to enhance spin fluctuations with a $\lambda\sim\mathcal{O}(U)$ necessary in order to stabilize the $BOW$ phase. We note that e-ph coupling scales inversely with phonon frequency ($\lambda=g^2/\omega$), so that this regime is within reach of material models. As discussed in Ref. \onlinecite{xing2021quantum}, the transition to the $BOW$ phase comes with a breaking of the symmetry of the square lattice; signatures of this breaking are present in the self-energy and are shown in Fig.\ref{qpSE}. Unlike the quasiparticle weight generated by the Holstein system, the SSH mode deforms the symmetric weight of the Hubbard model as the e-ph coupling is increased. Deep in the $BOW$ phase (rightmost panel), the symmetry of the nodal structure of the Hubbard model is lost and we see a quasiparticle weight consistent with a stable $BOW$ phase. The primary driver of this is the initial momentum structure of the phonon mode, with other phonon modes leading to possibly more exotic quasiparticle weights.

Beyond signatures in the electronic self-energy, the e-ph coupling enhances the phonon self-energy and spin susceptibility as we approach the transition to the $BOW$ phase. Fig.\ref{dopSusSSH} shows the response of the system doped away from half filling as the transition to $BOW$ is approached. The e-ph coupling enhances the incommensurate response with the $AF$ and $BOW$ susceptibilities showing peaks at an incommensurate wave vector. The interaction magnifies asymmetries between $(\pi,\pi-\delta q)$ vector and $(\pi-\delta q,\pi-\delta q)$ normally seen in the Hubbard model with $BOW$ ordering occurring firmly at the former vector for $p=0.18$. This preference is seen in the self-energy of the phonon with a clear trend towards the $(\pi,\pi)$ mode associated with the $BOW$ phase as the e-ph coupling is increased. The impact of doping on the $AF$ and $BOW$ response at the transition point between the phases is shown in Fig.\ref{dopPSusSSH}. Doping the electronic sector appears to have little impact on the self-energy of the phonon apart from moving its peak away from $(\pi,\pi)$. The impact is much larger on the $BOW$ response with incommensurate response seen for $p>0.15$. In this regime we see the appearance of a novel incommensurate $d_{x^2-y^2}$ charge response that dominate over the $BOW$ phase for $U=4t$. The charge response in the d-channel is shown in Fig.\ref{dChargeSSH}. Previous studies have considered possible $BOW$ configurations for the 2D SSH model with the $d_{x^2-y^2}$ and $p_{x/y}$ charge orders with the latter showing the larger energy gain at half filling\cite{tang1988peierls}. Monte Carlo studies confirmed the bond order for the half-filled SSH model but the impact of doping at finite $U$ remains unexplored\cite{xing2021quantum}. The vector at which we observe the d-type charge shows an asymmetric shift between the x and y directions, essentially aligning the bond charge of the $BOW$ phase into the plaquettes corresponding to $d_{x^2-y^2}$ order.

The phase diagram of the Extended Peierls Hubbard model is shown in Fig.\ref{phaseDiagSSH}. We find that a large enough e-ph coupling stabilizes the $BOW$ phase over the $AF$ phase for all values of $V$. The density-density interaction enhances charge order leading to an expansion of the $BOW$ for all values of the e-ph coupling. In the 1D system the interaction not only drives charge order but also stabilizes a finite $BOW$ phase that expands as function of $U$ in the absence of the e-ph interaction. This leads to a significant reduction of the $AF$ domain for the case of $V=U/2$. The behavior of the spin susceptibilities is similar in 1D and 2D with the system showing the expected enhancement as we approach the transition to $BOW$ followed by a suppression in the $BOW$ phase for all values of $V$.

\section{Conclusions}
\label{summary}

In this work, we have applied the decoupled fRG to study the phases in the Hubbard-Holstein and extended Peierls Hubbard Hamiltonians in one and two dimensions. The decoupled fRG allows for a computationally efficient inclusion of frequency modes in the vertex which are crucial for the study of the role played by interactions between the electrons and lattice. The fRG enables, given a phonon mode and electron-phonon vertex, access to the self-energies of the electron and phonon modes, at low temperatures ($\beta\sim 50$) for large system sizes. We account for the impact of these interactions on the response of the system by calculating the charge, spin and superconducting correlators. The fRG captures the various phases seen in these systems with different e-ph couplings, for different electron-electron interactions as function of phonon frequency, doping and temperature. Our results for the two e-ph systems considered here were cross-checked against DMRG and Monte Carlo studies in one and two dimensions. Despite limitations to moderate coupling, the success of the fRG indicates the possibility of addressing the impact of arbitrary phonon modes that couple locally to electronic Hamiltonians. The fRG also allows the study of systems with large phonon frequencies which are beyond the realm of material models. Such systems have been proposed in the cold atom setting, and the study of faster lattice dynamics on electronic orders can help paint a more complete picture of e-ph interactions\cite{hague2012quantum}.

We applied the two-loop fRG to the Hubbard-Holstein model with the two-pronged goal of validating the approach and exploring the impact of doping and phonon frequency on the two dimensional variants of the system. In one dimension, the fRG captures the metallic, $AF$ and $CDW$ phases in the system and reproduces the extension of the metallic phase to the Holstein model ($U=0$). The transition line shifted from $U=\lambda$ is reproduced with the extent of the metallic phase in line with previous DMRG and RG studies. In two dimensions we explored the impact of doping and phonon frequency on the metallic phase in the HH model. Our results for charge response of the system at various phonon frequencies and doping indicate a metallic phase that extends to the Holstein model ($U=0$) and expands with phonon frequency. The accompanying s-wave superconducting response grows with the e-ph coupling and doping though a finite $U$ coupling suppresses these correlations leaving just the d-type superconducting fluctuations usually seen in the Hubbard model.

For the Peierls-Hubbard model, we find a transition to a $BOW$ phase in one and two dimensions. The e-ph coupling in the model enhances $AF$ correlations with the system showing strong antiferromagnetic response even deep in the $BOW$ phase. As noted in previous works \cite{xing2021quantum}, the stabilization of the $BOW$ phase does break the symmetry of the square lattice. With the fRG we find evidence of this symmetry breaking in both electron and phonon self-energies. Doping the system leads to an incommensurate $BOW$ response which appears stable even at large e-ph coupling and shows sensitivity only to the frequency of the phonon. In this doped regime at moderate $U$ we found the incommensurate bond order switching to an $d_{x^2-y^2}$ charge order suggesting a change in the optimal bond ordering pattern for the doped regime. Inclusion of a nearest neighbor density density interaction shifts the transition line in favor of the $BOW$ phase. The magnitude of the shift appears much smaller in two dimensions, which suggests that it has little impact on the nature of the transition.

Our results for the two models suggest many further directions to explore, with the most rewarding possibly being the study of exotic phonons such as the $A_{1g}$ and $B_{1g}$ modes seen in the Cuprates. Given the two dimensional results, a more thorough study of the interplay between the phonon frequency, doping and e-ph coupling coupling for a three dimensional model should further clarify the nature of the metallic phase in the HH model. Work along these directions is currently in progress.
\section{Acknowledgment}
KMT is supported by NSF DMR-1728457. We wish to thank Shan-Wen Tsai for illuminating discussions during the course of this work. We would also like to thank Boston University's Research Computing Services for providing the computational resources required for the work.
\bibliography{ephFRG.bib}

\end{document}